\documentclass[usenatbib,fleqn]{mnras}

\usepackage{newtxtext,newtxmath}

\usepackage[T1]{fontenc}
\usepackage{ae,aecompl}

\usepackage{graphicx}	
\usepackage{amsmath}	
\usepackage{amssymb}	
\usepackage{ulem}

\usepackage{hyperref}

\usepackage{multirow}
\usepackage{array}  
\usepackage{booktabs}
\usepackage{times}
\usepackage{subfigure}

\usepackage[usenames,dvipsnames]{xcolor}

\title[Exploring accretion and jet in nearby NLS1s]{
Exploring the physics of the accretion and jet in nearby narrow-line Seyfert 1 galaxies
}

\author[S. Yao et al.]{
Su Yao$^{1}$\thanks{E-mail: yaosu@pku.edu.cn}, 
Erlin Qiao$^{2,3}$, 
Xue-Bing Wu$^{1,4}$, 
B. You$^{5,6}$
\\
$^{1}$Kavli Institute for Astronomy and Astrophysics, Peking University, Beijing 100871, China\\
$^{2}$National Astronomical Observatories, Chinese Academy of Sciences, Beijing 100012, China\\
$^{3}$School of Astronomy and Space Sciences, University of Chinese Academy of Sciences, 19A Yuquan Road, Beijing 100049, China\\
$^{4}$Department of Astronomy, School of Physics, Peking University, Beijing 100871, China\\
$^{5}$School of Physics and Technology, Wuhan University, Wuhan 430072, China\\
$^{6}$Copernicus Astronomical Center, Polish Academy of Science, Bartycka 18, 00-716 Warsaw, Poland
}

\date{Accepted XXX. Received YYY; in original form ZZZ}

\pubyear{2017}

\begin{document}
\label{firstpage}
\pagerange{\pageref{firstpage}--\pageref{lastpage}}
\maketitle

\begin{abstract}
In this paper, 
we explore the physics of the accretion and jet in narrow-line Seyfert 1 galaxies (NLS1). 
Specifically, we compile a sample composed of 16 nearby NLS1
with $L_{\rm bol}/L_{\rm Edd} \gtrsim 0.1$. 
We investigate the mutual correlation between their 
radio luminosity  $L_{\rm R}$,  X-ray luminosity $L_{\rm X}$,  
optical luminosity $L_{\rm 5100}$
and black hole mass  $M_{\rm BH}$. 
By adopting partial correlation analysis: 
(1) we find a positive correlation between $L_{\rm X}$ and $M_{\rm BH}$, 
and 
(2) we find a weak positive correlation between $L_{\rm R}$ and $L_{5100}$. 
However, 
we don't find significant correlations between $L_{\rm R}$ and $L_{\rm X}$ 
or between $L_{\rm X}$ and $L_{5100}$ 
after considering the effect of the black hole mass, 
which leads to a finding of the independence of $L_{\rm X}/L_{\rm Edd}$ on $L_{5100}/L_{\rm Edd}$.
Interestingly, the findings that $L_{\rm X}$ is correlated with $M_{\rm BH}$ 
and $L_{\rm X}/L_{\rm Edd}$ is not correlated with $L_{5100}/L_{\rm Edd}$ support that the X-ray emission is saturated with increasing $\dot{M}$ for $L_{\rm bol}/L_{\rm Edd} \gtrsim 0.1$ in NLS1s, 
which may be understood in the framework of slim disc  scenario.
Finally, we suggest that a larger NLS1 sample with  high quality radio and X-ray data is needed to further confirm this result in the future. 
\end{abstract}


\section{Introduction}
\label{intro}

Narrow-line Seyfert 1 galaxy (NLS1) is a subclass of broad-line active galactic nuclei (AGNs), 
identified optically by full width at half maximum (FWHM) of their broad H$\beta$ emission line 
$<2000\rm\,km\,s^{-1}$ 
and 
weak [O{\sc\,iii}] lines with [O{\sc\,iii}]$\rm\lambda5007/H\beta<3$ 
\citep[][]{1985ApJ...297..166O, 1989ApJ...342..224G}. 
Compared to normal Seyfert 1 galaxies, 
NLS1s show 
smaller $\rm FWHM(H\beta)$, 
weaker [O{\sc\,iii}], 
stronger Fe{\sc\,ii} emission, 
stronger blueshift of the [O{\sc\,iii}] and C{\sc\,iv}, 
steeper soft X-ray spectrum \citep[e.g.,][]{BG92, 2001A&A...372..730V, 2000ApJ...536L...5S, 2007ApJ...666..757S, 2004ApJ...611..107L, 1996A&A...309...81W,1996A&A...305...53B}. 
NLS1s are located at an extreme end of a correlation matrix of AGN observables, 
which is believed to be driven mainly by the Eddington ratio $L_{\rm bol}/L_{\rm Edd}$
\citep[e.g.,][]{BG92, 2002ApJ...565...78B}, 
where $L_{\rm bol}$ is the bolometric luminosity and $L_{\rm Edd}=1.3\times10^{38}M_{\rm BH}/M_{\odot}\rm\,erg\,s^{-1}$. 
Bulk of the black hole masses of the NLS1s has been measured to be $\sim10^{6-7}\,M_{\odot}$ by various methods [e.g.,the  reverberation mapping method 
\citep[e.g.,][]{2014ApJ...793..108W} 
and the luminosity-radius relation 
\citep[e.g.,][]{Zhou2006}]. 
They generally have high Eddington ratios close to or even above unity, 
so they are often suggested to be powered by slim discs \citep[e.g.,][]{1999ApJ...516..420W, 2000slim}.

NLS1s  are often being radio-quiet\footnote{
A measurement of the radio loudness is usually characterized by a parameter $R=f_{5\rm\,GHz}/f_{4400\,\AA}$, 
where $f_{5\rm\,GHz}$ and $f_{4400\,\AA}$ are the flux densities at $5\rm\,GHz$ and $4400$\,\AA. 
The optically selected AGNs can be conventionally classified as radio-quiet if $R\lesssim10$ \citep[][]{1989AJ.....98.1195K}. 
} 
compared with other broad-line AGNs 
\citep[e.g,][]{1995AJ....109...81U, 2003ApJ...588..746S, 2006AJ....131.1948W}. 
Only $\sim7\%$ of the optical selected NLS1s are radio-loud \citep[e.g.,][]{2006AJ....132..531K, Zhou2006}, 
while this fraction is $10\%-15\%$ for normal broad-line AGNs \citep{2002AJ....124.2364I}. 
By performing a comprehensive study of 23 genuine radio-loud NLS1s selected from a large NLS1 sample, 
\citet{2008ApJ...685..801Y} have found evidence of relativistic blazar-like jet in the radio loudest NLS1s. 
Then the powerful jet in the radio-loud NLS1s was confirmed by the $\gamma$-ray detections 
\citep[e.g.,][]{2009ApJ...699..976A, 2009ApJ...707L.142A, 2012MNRAS.426..317D, 2015MNRAS.454L..16Y}.
On the other hand, 
several works have found the evidence for the presence of 
jets in some radio-quiet NLS1s revealed by VLBI images, 
e.g., extended jet-like features and high brightness temperatures 
of the compact radio core component on pc scales
\citep[e.g.,][]{2004A&A...425...99L, 2009ApJ...706L.260G, 2013ApJ...765...69D, 2015ApJ...798L..30D}.

Generally, the X-ray emission in radio-quiet AGNs is believed to be produced in the hot corona by the inverse Compton scattering of the soft photons from the accretion disc 
\citep[e.g.,][]{1991ApJ...380L..51H, 1993ApJ...413..507H, 2002ApJ...575..117L}, 
while the origin of the radio emission in the radio-quiet AGNs is still not fully understood, which is suggested to be from jets by some works although not fully proved \citep[e.g.,][]{2016AN....337...73C}. 
It is shown that the study of the relation between the radio luminosity and X-ray luminosity
is a very useful tool to investigate the coupling of the accretion and the jet 
\citep{2003merloni}. 
The radio/X-ray correlation has been investigated for different types of AGN, as well as 
black hole X-ray binaries (BHBs) in the low/hard spectral state
\citep[e.g.,][]{2003merloni, 2006ApJ...645..890W, 
2008ApJ...688..826L, 
2011MNRAS.415.2910D}. 
It is found that $L_{5\rm\,GHz}\propto L_{2-10\rm\,keV}^{\sim0.5-0.7}$ in the low-luminosity AGNs (LLAGNs) and in the under-luminous low/hard spectral
state of BHBs 
\citep[e.g.,][]{2006A&A...456..439K}, 
while this correlation becomes significantly steeper, i.e., $L_{5\rm\,GHz}\propto L_{2-10\rm\,keV}^{\sim1.4}$, 
in the luminous AGNs and the luminous low/hard spectral state of BHBs
\citep[e.g.,][]{2004ApJ...617.1272C, 
2010A&A...524A..29R, 2014ApJ...787L..20D}

Some NLS1s have been included in previous studies of radio/X-ray correlation.
However, due to that the number of the NLS1s in the sample is very small, 
their statistical properties are overwhelmed by other types of AGNs. 
\citep[e.g.,][]{2003merloni, 2014ApJ...787L..20D, 2015MNRAS.447.1289P}.
Furthermore, the measurements of their radio luminosities could be severely contaminated by the host galaxies, 
e.g., if the resolution of the radio observation is not high enough, the star formation activities in their circumnuclear regions can have very strong contributions to the radio luminosity 
\citep[e.g.,][]{2006AJ....132..321D, 2007ApJS..169....1O, 2010MNRAS.403.1246S}. 
\citet{2015MNRAS.451.1795C} 
found that the star formation activities can contribute a significant fraction of VLA/FIRST radio emission even in some radio-loud NLS1s. 
In order to reduce the contaminations from the host galaxy as much as possible, 
in this work 
we make use of radio fluxes within the central regions ($\lesssim1\rm\,kpc$) of nearby radio-quiet NLS1s measured from high resolution radio images, 
to study their nuclear radio/X-ray relation 
which was rarely studied before. 
This could help us understand the physics of the accretion and jet in high Eddington ratio regime. 
We also explore the mutual relation between the radio luminosity, X-ray luminosity, optical luminosity and black hole mass. 
There are several works on the radio images of 
radio loud NLS1s on kpc scales or even pc scales recently \citep[e.g.,][]{2011ApJ...738..126D, 2012ApJ...760...41D, 2015ApJS..221....3G}. 
But we do not choose the radio loud ones as to avoid the relativistically beaming effects. 
Throughout this work a cosmology is assumed with $H_0=70$ km s$^{-1}$ Mpc$^{-1}$, $\Omega_\Lambda=0.73$ and $\Omega_{\rm M}=0.27$. 
The luminosities taken from literatures would be transformed to the cosmology used here when necessary. 

\section{The Sample}
\label{sample}

\subsection{Sample selection}

We select objects from the 13th edition of the catalogue of quasars and active galactic nuclei compiled by \citet{2010A&A...518A..10V}, 
including both quasars (defined as brighter than absolute magnitude $M_{B}=-22.25$) 
and Seyfert galaxies (fainter than $M_{B}=-22.25$), with criteria as follows. 
\begin{enumerate}
	\item The object is classified as NLS1 with redshift $z<0.1$ according to its optical spectroscopy. 
        \item The object is radio-quiet. 
	\item The object is detected by the high-resolution radio observations within the central nuclear region less than about 1 kpc. 
	\item The object has an Eddington ratio of $L_{\rm bol}/L_{\rm Edd}\gtrsim 0.1$. 
\end{enumerate}

The criterion of being classified as NLS1 in \citet{2010A&A...518A..10V} 
is that the  FWHM of the H$\beta$ emission line broad component is narrower than $2000\rm\,km\,s^{-1}$. 
265 NLS1s are picked out at this step. 
Then, we search for the corresponding radio observations by VLA of the selected NLS1s 
with an angular resolution corresponding to a physical size of $\lesssim1\rm\,kpc$ at the object's rest frame from the literatures, i.e., 
\citet{1984ApJ...285..439U}, 
\citet{1993MNRAS.263..425M}, 
\citet{1995MNRAS.276.1262K}, 
\citet{1995AJ....109...81U}, 
\citet{1998MNRAS.297..366K}, 
\citet{2000ApJ...537..152K}, 
\citet{2001ApJS..132..199S}, 
\citet{2001ApJS..133...77H}, 
\citet{2004A&A...425...99L}, 
\citet{2010MNRAS.401.2599O} 
and 
\citet{2010ApJ...720..555P}. 
After this step, we have 17 sources left. 

We estimate the Eddington ratios for these 17 sources (see section~\ref{bhmass} for details). 
All of them have $L_{\rm bol}/L_{\rm Edd}>0.1$ except for NGC~4051. 
The nuclear continuum flux at 5100\,\AA~ of NGC~4051 is about $5\times10^{-15}\rm\,erg\,s^{-1}\,cm^{-2}$ after subtracting the host galaxy starlight 
and its black hole mass estimated from the reverberation mapping method is $1.73\times10^{6}\,M_{\odot}$ \citep{2009ApJ...702.1353D}. 
The Eddington ratio is $\sim1\%$ for NGC~4051.
Thus we exclude it from our sample. 
The final sample consists of 16 NLS1s with  $z<0.1$ and  $L_{\rm bol}/L_{\rm Edd}>0.1$ (Table~\ref{rqnls1_sample}). 
None of them are reported to be radio-loud in previous works 
\citep[e.g.,][]{2006AJ....132..531K, 2008ApJ...685..801Y, 2015A&A...575A..13F}. 
In the work of \citet{2015A&A...578A..28B}, 
the radio loudness of Mrk~1239 and Mrk~766 have been reported as 16 and 23, respectively. 
They could be classified as so-called radio-intermediate AGNs instead of genuine radio-loud ones \citep{1996ApJ...471..106F}, 
which are also included in our sample. 
Here we should note that this sample is not complete, 
since we only select sources that are detected in high resolution radio observations. 
But this is the largest sample we can find from the literatures at this stage.

\newcounter{tabref}
\begin{table*}
	\caption{Radio, X-ray and optical luminosities of the radio-quiet NLS1 sample with the measured black hole masses. }
	\label{rqnls1_sample}
	\begin{center}
	\setlength{\tabcolsep}{3.0pt} 
	\begin{tabular}{lcccccccclccclc}
		\hline
		 Name & $z$ & $\log L_{\rm R}$ & size & Ref. & $\log L_{\rm R}$ & size & Ref. & $\log L_{\rm X}$ & Ref. & $\log L_{\rm HX}$ & $\log M_{\rm BH}$ & $\log L_{5100}$ & Ref. & $\lambda$ \\ 
		         &           &  VLA  &  & &  VLBI  & & & $2-10\rm\,keV$ & & $14-195\rm\,keV$ &  \\
		& & (ergs\,s$^{-1}$) & (pc) & & (ergs\,s$^{-1}$) & (pc) & & (ergs\,s$^{-1}$) & & (ergs\,s$^{-1}$) & ($M_{\odot}$) & (ergs\,s$^{-1}$) \\
		 (1)  &  (2)  &  (3)  &  (4)  &  (5)  &  (6)  &  (7)  &  (8)  &  (9)  &  (10)  &  (11)  & (12) & (13) & (14) & (15) \\ 
		\hline
		Mrk 335 & 0.0258 & 38.31 & 152 & \ref{Kukula1995} & - & - & - & 43.17 & \ref{geo2000},\ref{grupe2007},\ref{grupe2010},\ref{keek2016} & 43.45 & 6.92 & 43.60 & \ref{Wang2014},M & 0.38 \\
		I Zw 1 & 0.0611 & 38.71 & 385 & \ref{Kukula1995} & - & - & - & 43.72 & \ref{Pico2005},\ref{Reeves2000},\ref{Costa2007},\ref{ueda2001} & - & 7.26 & 44.51 & \ref{Wang2013},S & 1.40 \\
		Mrk 359 & 0.0168 & 37.36 & 170 & \ref{Kinney2000} & - & - & - & 42.60 & \ref{lutz2004},\ref{bianchi2009},\ref{boissay2016} & 42.93 & 5.48 & 41.99$^{b}$ & \ref{ho2008},S & 0.25 \\
		Mrk 110 & 0.0353 & 38.42 & 320 & \ref{Miller1993} & 37.86 & 7.5 & \ref{Doi2013} & 43.83 & \ref{geo2000},\ref{nandra2007},\ref{winter2012} & 44.22 & 7.05 & 43.61 & \ref{Wang2014},M & 0.28 \\
		Mrk 705 & 0.0292 & 38.45 & $<55$ & \ref{Schmitt2001} & 37.91 & 6.5 & \ref{Doi2013} & 43.48 & \ref{gallo2005},\ref{shu2010} & 43.49 & 6.79 & 43.04$^{b}$ & \ref{ho2008},\ref{ma2003},S & 0.14 \\
		Mrk 1239 & 0.0199 & 38.93 & 140 & \ref{Ulvestad1995} & 37.70 & 4.5 & \ref{Doi2013} & 43.02 & \ref{grupe2004} & 43.05$^{a}$ & 6.11 & 43.27 & \ref{ryan2007},S & 1.12 \\
		\multirow{2}{*}{Mrk 766} & \multirow{2}{*}{0.0129} & \multirow{2}{*}{38.41} & \multirow{2}{*}{95} & \multirow{2}{*}{\ref{Parra2010}} & \multirow{2}{*}{37.04} & \multirow{2}{*}{2.9} & \multirow{2}{*}{\ref{Doi2013}} & \multirow{2}{*}{42.84} & \ref{nandra2007},\ref{shu2010},\ref{matt2000}, & \multirow{2}{*}{42.91} & \multirow{2}{*}{6.11} & \multirow{2}{*}{42.53} & \multirow{2}{*}{\ref{grier2013},M} & \multirow{2}{*}{0.20} \\
		& & & & & & & & & \ref{ueda2001},\ref{landi2005},\ref{Giacche2014} & & \\
		PG 1244+026 & 0.0481 & 38.13 & 434 & \ref{Miller1993} & - & - & - & 43.17 & \ref{Pico2005},\ref{ueda2001},\ref{Jin2013} & - & 6.28 & 43.43 & \ref{Wang2013},S & 1.11 \\
		Mrk 783 & 0.0672 & 39.27 & $<380$ & \ref{Ulvestad1984a} & 38.54 & 13.6 & \ref{Doi2013} & 43.84 & \ref{Panessa2011} & 44.28 & 7.16 & 43.91$^{a}$ & \ref{wang2001},S & 0.44 \\
		PG 1404+226 & 0.0983 & 38.95 & 837 & \ref{Miller1993} & - & - & - & 42.88 & \ref{Pico2005},\ref{ueda2001} & - & 6.89 & 44.38 & \ref{vest2006},S & 2.40 \\
		Mrk 478 & 0.0790 & 38.76 & 688 & \ref{Miller1993} & - & - & - & 43.84 & \ref{Pico2005},\ref{ueda2001},\ref{shinozaki2006},\ref{inoue2007} & - & 7.37 & 44.48 & \ref{Wang2013},S & 1.02 \\
		PG 1448+273 & 0.0645 & 38.76 & 571 & \ref{Miller1993} & - & - & - & 43.33 & \ref{inoue2007} & - & 7.01 & 44.01 & \ref{Wang2013},S & 0.77 \\
		IRAS 1509$-$211 & 0.0445 & 39.44 & $<830$ & \ref{Ulvestad1995} & - & - & - & 43.57 & \ref{brightman2011},\ref{Liu2014},\ref{Liu2015} & 44.16 & 7.00 & 43.98 & \ref{Ohta2007},S & 0.74 \\ 
		Mrk 493 & 0.0310 & 37.92 & $<260$ & \ref{Ulvestad1995} & - & - & - & 42.95 & \ref{bianchi2009} & - & 6.18 & 43.03 & \ref{Wang2014},M & 0.55 \\
		Mrk 507 & 0.0559 & 38.70 & $<1320$ & \ref{Ulvestad1995} & - & - & - & 42.81 & \ref{bianchi2009},\ref{iwasawa1998},\ref{de2010} & 43.56 & 6.91 & 43.77 & \ref{Ohta2007},S & 0.56 \\
		\multirow{2}{*}{Ark 564} & \multirow{2}{*}{0.0247} & \multirow{2}{*}{38.93} & \multirow{2}{*}{320} & \multirow{2}{*}{\ref{Lai2004}} & \multirow{2}{*}{38.40} & \multirow{2}{*}{1.2} & \multirow{2}{*}{\ref{Lai2004}} & \multirow{2}{*}{43.36} & \ref{nandra2007},\ref{shinozaki2006},\ref{turner2001},\ref{vignali2004}, & \multirow{2}{*}{43.52$^{a}$} & \multirow{2}{*}{6.42} & \multirow{2}{*}{43.68} & \multirow{2}{*}{\ref{botte2004},S} & \multirow{2}{*}{1.44} \\
		& & & & & & & & & \ref{matsumoto2004},\ref{papadakis2007},\ref{ram2013} & & \\
		\hline
	\end{tabular}
	\parbox[]{\textwidth}{
	$^a$ This luminosity is converted from $10-50\rm\,keV$ measured by {\it Suzaku}/HXD \citep{2011ApJ...727...19F} assuming a photon index of $\Gamma=2$. \\
	$^b$ This monochromatic luminosity is estimated from $L_{\rm H\beta}-L_{5100}$ relation of \citet{2005ApJ...630..122G} and $L_{\rm H\beta}$ is obtained from the corresponding references in Column (14). 
	\\
	{\it Note.} 
	Column (1): source name. 
	Column (2): source redshift. 
	Column (3): the VLA monochromatic radio luminosity at 5\,GHz. 
	Column (4): the linear size of the emitting region from which Column (3) is measured. 
	Column (5): the references for Column (3) and (4). 
	Column (6): the VLBI monochromatic radio luminosity at 5\,GHz, 
	Column (7): the linear size of the emitting region from which Column (6) is measured. 
	Column (8): the references for Column (6) and (7). 
	Column (9): the average X-ray luminosity in the $2-10\rm\,keV$ band. 
	Column (10): the references for Column (9). 
	Column (11): the hard X-ray luminosity in the $14-195\rm\,keV$ band taken from the 70 month {\it Swift}/BAT survey \citep{2013ApJS..207...19B}, except for Mrk 1239 and Ark 564, for which the hard X-ray luminosity in the $10-50\rm\,keV$ was taken from \citet{2011ApJ...727...19F}, and then converted to the $14-195\rm\,keV$ band. 
	Column (12): the black hole masses. 
	Column (13): monochromatic nuclear luminosity at $5100$\,\AA. 
	Column (14): references for the $M_{\rm BH}$ and $L_{5100}$. The black hole masses are derived from the reverberation mapping method (M) or single-epoch spectroscopy (S). 
	Column (15): Eddington ratio.
	\\
	{\it References:} 
	\refstepcounter{tabref}\label{Kukula1995}(\ref{Kukula1995}) \citet{1995MNRAS.276.1262K}; 		\refstepcounter{tabref}\label{Kinney2000}(\ref{Kinney2000}) \citet{2000ApJ...537..152K}; 
	\refstepcounter{tabref}\label{Miller1993}(\ref{Miller1993}) \citet{1993MNRAS.263..425M}; 
	\refstepcounter{tabref}\label{Schmitt2001}(\ref{Schmitt2001}) \citet{2001ApJS..132..199S}; 
	\refstepcounter{tabref}\label{Ulvestad1995}(\ref{Ulvestad1995}) \citet{1995AJ....109...81U}; 
	\refstepcounter{tabref}\label{Parra2010}(\ref{Parra2010}) \citet{2010ApJ...720..555P}; 
	\refstepcounter{tabref}\label{Ulvestad1984a}(\ref{Ulvestad1984a}) \citet{1984ApJ...278..544U}; 
	\refstepcounter{tabref}\label{Lai2004}(\ref{Lai2004}) \citet{2004A&A...425...99L}; 
	\refstepcounter{tabref}\label{Doi2013}(\ref{Doi2013}) \citet{2013ApJ...765...69D}; 
	\refstepcounter{tabref}\label{geo2000}(\ref{geo2000}) \citet{2000ApJ...531...52G}; 
	\refstepcounter{tabref}\label{grupe2007}(\ref{grupe2007}) \citet{2007ApJ...668L.111G}; 
	\refstepcounter{tabref}\label{grupe2010}(\ref{grupe2010}) \citet{2010ApJS..187...64G}; 
	\refstepcounter{tabref}\label{keek2016}(\ref{keek2016}) \citet{2016MNRAS.456.2722K}; 
	\refstepcounter{tabref}\label{Pico2005}(\ref{Pico2005}) \citet{2005A&A...432...15P};
	\refstepcounter{tabref}\label{Reeves2000}(\ref{Reeves2000}) \citet{2000MNRAS.316..234R};
	\refstepcounter{tabref}\label{Costa2007}(\ref{Costa2007}) \citet{2007MNRAS.378..873C};
	\refstepcounter{tabref}\label{ueda2001}(\ref{ueda2001}) \citet{2001ApJS..133....1U}; 
	\refstepcounter{tabref}\label{lutz2004}(\ref{lutz2004}) \citet{2004A&A...418..465L}; 
	\refstepcounter{tabref}\label{bianchi2009}(\ref{bianchi2009}) \citet{2009A&A...495..421B}; 
	\refstepcounter{tabref}\label{boissay2016}(\ref{boissay2016}) \citet{2016A&A...588A..70B}; 
	\refstepcounter{tabref}\label{nandra2007}(\ref{nandra2007}) \citet{2007MNRAS.382..194N}; 
	\refstepcounter{tabref}\label{winter2012}(\ref{winter2012}) \citet{2012ApJ...745..107W}; 
	\refstepcounter{tabref}\label{gallo2005}(\ref{gallo2005}) \citet{2005A&A...442..909G}; 
	\refstepcounter{tabref}\label{shu2010}(\ref{shu2010}) \citet{2010ApJS..187..581S}; 
	\refstepcounter{tabref}\label{grupe2004}(\ref{grupe2004}) \citet{2004AJ....127.3161G}; 
	\refstepcounter{tabref}\label{matt2000}(\ref{matt2000}) \citet{2000A&A...363..863M}; 
	\refstepcounter{tabref}\label{landi2005}(\ref{landi2005}) \citet{2005A&A...441...69L}; 
	\refstepcounter{tabref}\label{Giacche2014}(\ref{Giacche2014}) \citet{2014A&A...562A..44G}; 
	\refstepcounter{tabref}\label{Jin2013}(\ref{Jin2013}) \citet{2013MNRAS.436.3173J}; 
	\refstepcounter{tabref}\label{Panessa2011}(\ref{Panessa2011}) \citet{2011MNRAS.417.2426P}; 
	\refstepcounter{tabref}\label{shinozaki2006}(\ref{shinozaki2006}) \citet{2006AJ....131.2843S}; 
	\refstepcounter{tabref}\label{inoue2007}(\ref{inoue2007}) \citet{2007ApJ...662..860I}; 
	\refstepcounter{tabref}\label{brightman2011}(\ref{brightman2011}) \citet{2011MNRAS.413.1206B}; 
	\refstepcounter{tabref}\label{Liu2014}(\ref{Liu2014}) \citet{2014ApJ...783..106L}; 
	\refstepcounter{tabref}\label{Liu2015}(\ref{Liu2015}) \citet{2015MNRAS.447..517L}; 
	\refstepcounter{tabref}\label{iwasawa1998}(\ref{iwasawa1998}) \citet{1998MNRAS.293..251I}; 
	\refstepcounter{tabref}\label{de2010}(\ref{de2010}) \citet{2010A&A...524A..50D}; 
	\refstepcounter{tabref}\label{turner2001}(\ref{turner2001}) \citet{2001ApJ...561..131T}; 
	\refstepcounter{tabref}\label{vignali2004}(\ref{vignali2004}) \citet{2004MNRAS.347..854V}; 
	\refstepcounter{tabref}\label{matsumoto2004}(\ref{matsumoto2004}) \citet{2004ApJ...603..456M}; 
	\refstepcounter{tabref}\label{papadakis2007}(\ref{papadakis2007}) \citet{2007A&A...461..931P}; 
	\refstepcounter{tabref}\label{ram2013}(\ref{ram2013}) \citet{2013A&A...551A..95R}; 
	\refstepcounter{tabref}\label{Wang2014}(\ref{Wang2014}) \citet{2014ApJ...793..108W}; 
	\refstepcounter{tabref}\label{Wang2013}(\ref{Wang2013}) \citet{2013PhRvL.110h1301W}; 
	\refstepcounter{tabref}\label{ho2008}(\ref{ho2008}) \citet{2008ApJS..177..103H}; 
	\refstepcounter{tabref}\label{ma2003}(\ref{ma2003}) \citet{2003ApJS..145..199M}; 
	\refstepcounter{tabref}\label{ryan2007}(\ref{ryan2007}) \citet{2007ApJ...654..799R}; 
	\refstepcounter{tabref}\label{grier2013}(\ref{grier2013}) \citet{2013ApJ...773...90G}; 
	\refstepcounter{tabref}\label{wang2001}(\ref{wang2001}) \citet{2001A&A...377...52W}; 
	\refstepcounter{tabref}\label{vest2006}(\ref{vest2006}) \citet{2006ApJ...641..689V}; 
	\refstepcounter{tabref}\label{Ohta2007}(\ref{Ohta2007}) \citet{2007ApJS..169....1O}; 
	\refstepcounter{tabref}\label{botte2004}(\ref{botte2004}) \citet{2004AJ....127.3168B}; 
	}
	\end{center}
\end{table*}

\subsection{Radio luminosity}

We collect the radio flux for each source from the references listed in Column (5) of Table~\ref{rqnls1_sample}. 
They usually have steep radio spectra ($\alpha_{\rm r}<-0.5$, where $S_{\nu}\propto\nu^{\alpha_{\rm r}}$), not remarkably different from the normal Seyfert 1 galaxies \citep{1993MNRAS.263..425M, 1995AJ....109...81U}. 
As shown by the radio images given in these references, 
most of the sources have compact and unresolved morphology within a few hundred pc in central region.
A few exceptions exist, e.g.,
Mrk 110 and Mrk 1239 show extended features on kpc-scales \citep{1993MNRAS.263..425M, 1994AJ....108.1163K, 2015ApJ...798L..30D},
and Ark 564 has a triple radio component along the north-south direction \citep{2001MNRAS.325..737T}.

We calculate the core radio luminosities from their peak flux densities of the core component.
The observations were performed at different frequencies.
We always use the flux density observed at $\sim5\rm\,GHz$ if available.
When the observation was taken at other frequencies, 
the flux is extrapolated from the observing center frequency to 5\,GHz 
using the radio spectral index if available or using an typical spectral index of $\alpha_{\rm r}=-0.7$ \citep[e.g.,][]{2013ApJ...765...69D}. 
The luminosities are listed in Table~\ref{rqnls1_sample}. 
A typical value of $\sigma_{\rm R}=0.2$ dex \citep[][]{2001ApJ...555..650H} is adopted as the uncertainty of the radio luminosity. 

Six of these objects, namely, Mrk 110, Mrk 705, Mrk 783, Mrk 766 \citep{2013ApJ...765...69D}, Mrk 1239 \citep{2013ApJ...765...69D, 2015ApJ...798L..30D} and Ark 564 \citep{2004A&A...425...99L}, 
have also been detected by VLBI observations and exhibit radio core with angular size of a few milliarcsecond, 
corresponding to a linear size of only a few pc. 
The sources show clear elongated features extending to one side or both sides of the core with a brightness peak. 
In addition, their core components have high brightness temperatures of $T_{\rm B}\gtrsim10^{7}\rm\,K$ \citep[][]{2013ApJ...765...69D, 2004A&A...425...99L}. 
The VLBI luminosities of the sources are also listed in the table.

\subsection{X-ray luminosity}

For the X-ray luminosities, we have searched NASA/IPAC Extragalactic Database (NED) and  literatures in which the 2--10\,keV 
fluxes or luminosities of the selected NLS1s are reported (Column 10 in Table~\ref{rqnls1_sample}). 
The selected objects are all nearby with $z<0.1$ and relatively bright in the X-rays. 
The central AGN is generally 2--3 orders of magnitude more luminous than the typical level of 
X-ray sources of the host galaxy, e.g., X-ray binaries and supernova remnants \citep{1989ARA&A..27...87F}. 

The fluxes are corrected for the Galactic absorption before converted to luminosities. 
For most of the sources in the sample, 
the spectral analysis results given in the literatures show weak intrinsic absorption 
with $N_{\rm H}<10^{22}\rm\,cm^{-2}$, 
which consequently will not significantly affect the hard X-ray flux in our study. 
\citet{2005A&A...432...15P} reported an intrinsic gas column density of $N_{\rm H}=1.4\times10^{22}\rm\,cm^{-2}$ for PG~1404+226 from the spectral fit, 
which will lead to an underestimation of the $2-10\rm\,keV$ luminosity by $\sim16$ percent for a spectrum with $\Gamma=2$. 
Another peculiar object worth mentioning is Mrk 507. 
This source has a very strong Fe{\sc\,ii} with $R_{4570}=1.94$ \citep{2001A&A...372..730V}, 
on the other hand, it exhibits a flat {\it ROSAT} $0.1-2.4\rm\,keV$ spectrum with $\Gamma=1.6$ 
which was suggested to be caused by intrinsic absorption with a column density of $N_{\rm H}=(2-3)\times10^{21}\rm\,cm^{-2}$ \citep{1998MNRAS.293..251I}. 
Its {\it ROSAT} spectrum is still flat ($\Gamma\sim2.4$) even after correction of intrinsic absorption \citep{1998MNRAS.293..251I}. 
With a extremely strong Fe{\sc\,ii} and the flattest soft X-ray spectrum, Mrk 507 is unusual in the light of positive correlation between the Fe{\sc\,ii} strength and the steepness of {\it ROSAT} spectrum \citep{1996A&A...309...81W}. 
However, 
the suggested intrinsic absorption should have little affection (a few percent) to the X-rays above $2\rm\,keV$. 
Nevertheless, we adopt the flux reported in literatures after correction of only Galactic absorption for all the sources.

We note here that the X-ray variability is very common in AGNs and its variability amplitude usually ranges from a few times 
to more than an order of magnitude on timescales of months up to years \citep[e.g.,][]{1993ARA&A..31..717M}. 
The measurement of the X-ray and other observables at different epochs may lead to large uncertainties when studying their relation. 
So we take the unweighted mean X-ray flux when multiple observations are available for each source. 
Although there are significant X-ray variabilities between different X-ray observations for individual sources, 
they varied within an order of magnitude. 
The photon indices of the $2-10\rm\,keV$ spectra of the sample are in the range of $\Gamma\sim1.6-2.9$ 
and do not varied much either ($\Delta\Gamma\lesssim0.5$) for individual sources\footnote{\citet{2014A&A...562A..44G} have analyzed {\it XMM-Newton} observations on Mrk~766 between 2000 and 2005, and find the flux $F_{2-10\rm\,keV}$ varied by a factor of $\sim5$ with photon index between 0.95 and 2.1. 
But they suggest the X-ray emission is probably affected by occultations during $\Gamma<1.6$. 
So we do not adopt the fluxes with $\Gamma<1.6$. }. 
Four of these objects, 
Mrk 110, Mrk 335, Mrk 705 
and Ark 564 have been monitored by {\it RXTE}/ASM
and/or by {\it MAXI}. 
We check the ASM and MAXI 1-day-binned light curves, 
and find that the sources have not been detected for most of the time. 
Only in few cases the light curve demonstrated flares where the fluxes increase possibly by more than an order of magnitude. 
The value of 2--10\,keV fluxes of each source we adopted here do not vary as large as during the flares, 
so they were most probably in their typical flux state during the observations. 
We adopt the range between highest and lowest values as the uncertainty for each source if there are multiple X-ray observations. 
If there is only one X-ray observation, 
a typical uncertainty of $\sigma_{\rm X}=0.23$\,dex is adopted \citep[][]{2005AJ....130..387S}.

Eight sources in the sample are reported in the Swift-BAT 70-Month catalog \citep{2013ApJS..207...19B}. 
For another two sources such as Mrk 1239 and Ark 564, X-ray luminosities are converted from $10-50\rm\,keV$ measured by {\it Suzaku}/HXD \citep{2011ApJ...727...19F} to $14-195\rm\,keV$ assuming a photon index of $\Gamma=2$ and $\sigma_{\rm X}=0.23$\,dex for these two sources 
(see Column 11 of Table~\ref{rqnls1_sample}). 

\subsection{Black hole mass and Eddington ratio}
\label{bhmass}
Some of the NLS1s in the sample are included in large reverberation-mapping campaigns 
\citep[e.g.,][]{
2013ApJ...773...90G, 2014ApJ...793..108W}, 
in which the black hole masses are estimated by measuring the structure of the broad-line region using mapping technique. 
For the rest of the sources, their black hole masses are estimated with the line width-luminosity mass scaling relation using single-epoch optical spectroscopy (see column 14 of Table~\ref{rqnls1_sample}). 
The black hole masses are in the range of $10^{5.5-7.4}\,M_{\odot}$, 
as the typical value of NLS1 population 
\citep{Zhou2006}.

The Eddington ratio of the NLS1s studied here were estimated in numerous works \citep[e.g.,][]{2014ApJ...793..108W, 2007ApJS..169....1O, 2003ApJS..145..199M, 2008ApJS..177..103H, 2013ApJ...773...90G, 2001A&A...377...52W, 2007ApJ...654..799R, 2004AJ....127.3168B}. 
All of the results support that these sources are accreting close to or above Eddington limit. 
We simply estimate their Eddington ratios $\lambda=L_{\rm bol}/L_{\rm Edd}$ (Column 15 in Table~\ref{rqnls1_sample}) assuming $L_{\rm bol}=9.8L_{5100}$ \citep{2004MNRAS.352.1390M}, 
where $L_{5100}$ is the monochromatic luminosities at $5100$\,\AA. 
For most of the sources in the sample, 
the $L_{5100}$ are calculated from the nuclear monochromatic fluxes at 5100\,\AA~after subtracting the starlight of the host galaxy (see references in Column 14 of Table~\ref{rqnls1_sample}). 
When the starlight-subtracted flux at $5100$\,\AA~is unavailable, we calculate $L_{5100}$ using the measured $L_{\rm H\beta}$ and the $L_{\rm H\beta}-L_{5100}$ relation in \citet{2005ApJ...630..122G}. 
We note that the bolometric correction factor of $9.8$ adopted here was measured based on a large sample of SDSS quasars \citep[][]{2004MNRAS.352.1390M}.

\section{Correlation Analysis}
\label{correlations}

\begin{figure*}
\centering
	\begin{tabular}{cc}
		\includegraphics[width=0.90\columnwidth]{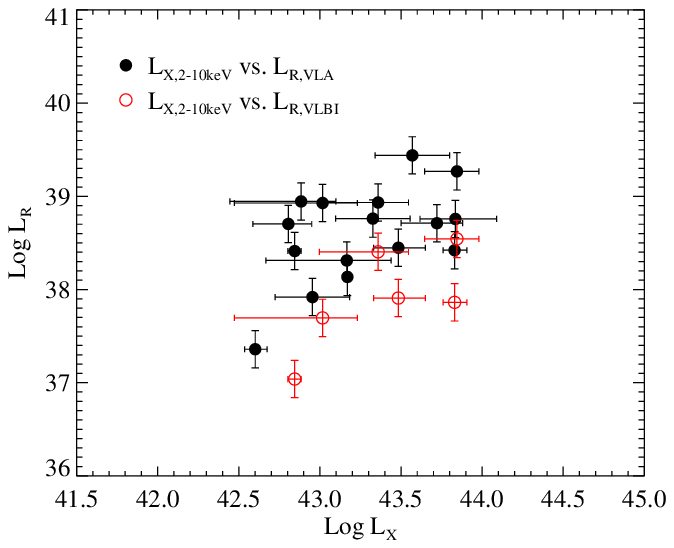} &
		\includegraphics[width=0.90\columnwidth]{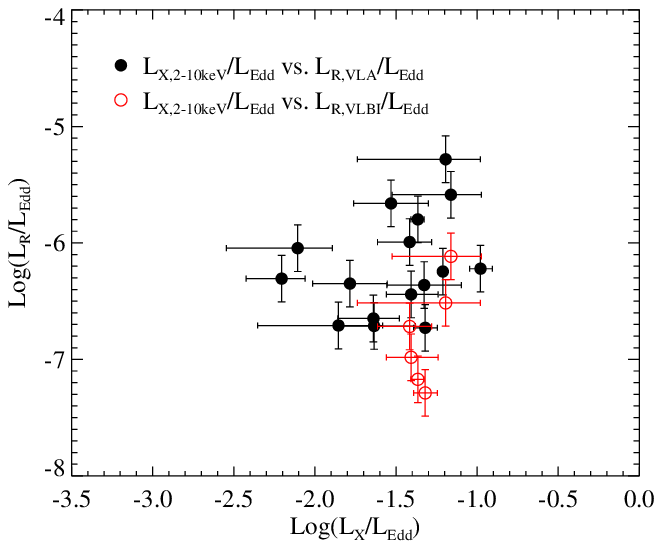} \\
		\includegraphics[width=0.90\columnwidth]{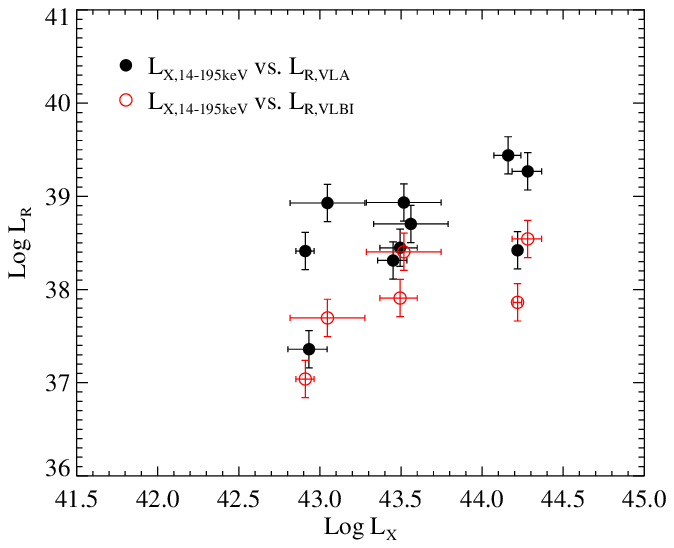} &
		\includegraphics[width=0.90\columnwidth]{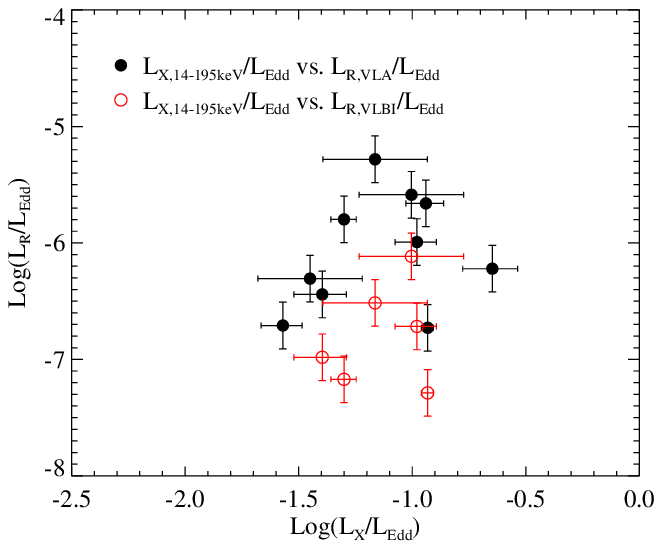} \\
	\end{tabular}
	\caption{
	The $L_{\rm X,2-10\,keV}$ versus $L_{\rm R}$ ({\it upper left}), 
	$L_{\rm X,2-10\,keV}/L_{\rm Edd}$ versus $L_{\rm R}/L_{\rm Edd}$ ({\it upper right}), 
	$L_{\rm X,14-195\,keV}$ versus $L_{\rm R}$ ({\it lower left}) 
	and 
	$L_{\rm X,14-195\,keV}/L_{\rm Edd}$ versus $L_{\rm R}/L_{\rm Edd}$ ({\it lower right}). 
	The VLA and VLBI data are represented by black filled circles and red open circles respectively. 
	}
\label{xvr}
\end{figure*}

\begin{table}
\renewcommand{\arraystretch}{1.2}
	\caption{Results of correlation analysis. }
	\label{cor_tab}
	\begin{center}
	\setlength{\tabcolsep}{3.0pt} 
	\begin{tabular}{ccccccccc}
		\hline
		\multicolumn{3}{c}{Variables} &   \multicolumn{2}{c}{Correlation} \\ 
		$X$ & $Y$ & $Z$ &  $\tau$ & $P_{\rm null}$ \\ 
		(1)  &  (2)  &  (3)  & (4)  &  (5)  \\ 
		\hline
		$\log L_{\rm X,2-10keV}$ & $\log L_{\rm R,VLA}$ & $\log M_{\rm BH}$ & 0.14 & 0.47 \\
		$\log L_{\rm X,2-10keV}$ & $\log L_{\rm R,VLBI}$ & $\log M_{\rm BH}$ & 0.31 & 0.45 \\
		$\log L_{\rm X,14-195keV}$ & $\log L_{\rm R,VLA}$ & $\log M_{\rm BH}$ & 0.37 & 0.17 \\
		$\log L_{\rm X,14-195keV}$ & $\log L_{\rm R,VLBI}$ &$\log M_{\rm BH}$ & 0.59 & 0.15 \\
		$\log L_{\rm R,VLA}$ & $\log L_{5100}$ & $\log M_{\rm BH}$ & 0.39 & 0.04 \\
		$\log L_{\rm R,VLBI}$ & $\log L_{5100}$ & $\log M_{\rm BH}$ & 0.62 & 0.13 \\
		$\log L_{\rm X,2-10keV}$ & $\log L_{5100}$ & $\log M_{\rm BH}$ & -0.02 & 0.94 \\
		$\log L_{\rm X,14-195keV}$ & $\log L_{5100}$ & $\log M_{\rm BH}$ & 0.42 & 0.12 \\
		$\log L_{\rm X,2-10keV}$ & $\log L_{\rm R,VLA}$ & $\log L_{5100}$ & 0.14 & 0.47  \\
		$\log L_{\rm X,2-10keV}$ & $\log L_{\rm R,VLBI}$ & $\log L_{5100}$ & 0.29 & 0.47 \\
		$\log L_{\rm X,14-195keV}$ & $\log L_{\rm R,VLA}$ & $\log L_{5100}$ & -0.02 & 0.95 \\
		$\log L_{\rm X,14-195keV}$ & $\log L_{\rm R,VLBI}$ & $\log L_{5100}$ & 0.42 & 0.30 \\
		$\log L_{\rm R,VLA}$ & $\log M_{\rm BH}$ & $\log L_{5100}$ & 0.01 & 0.97 \\
		$\log L_{\rm R,VLBI}$ & $\log M_{\rm BH}$ & $\log L_{5100}$ & 0.26 & 0.52 \\
		$\log L_{\rm R,VLA}$ & $\log M_{\rm BH}$ & $\log L_{\rm X,2-10keV}$ & 0.20 & 0.31 \\
		$\log L_{\rm R,VLBI}$ & $\log M_{\rm BH}$ & $\log L_{\rm X,2-10keV}$ & -0.13 & 0.74 \\
		$\log L_{\rm R,VLA}$ & $\log M_{\rm BH}$ & $\log L_{\rm X,14-195keV}$ & -0.07 & 0.78 \\
		$\log L_{\rm R,VLBI}$ & $\log M_{\rm BH}$ & $\log L_{\rm X,14-195keV}$ & -0.14 & 0.72 \\
		$\log L_{\rm X,2-10keV}$ & $\log M_{\rm BH}$ & $\log L_{5100}$ & 0.52 & $7.28\times10^{-3}$ \\
		$\log L_{\rm X,2-10keV}$ & $\log M_{\rm BH}$ & $\log L_{\rm R,VLA}$ & 0.58 & $2.47\times10^{-3}$ \\
		$\log L_{\rm X,2-10keV}$ & $\log M_{\rm BH}$ & $\log L_{\rm R,VLBI}$ & 0.95 & 0.02 \\
		$\log L_{\rm X,14-195keV}$ & $\log M_{\rm BH}$ & $\log L_{5100}$ & 0.59 & 0.03 \\
		$\log L_{\rm X,14-195keV}$ & $\log M_{\rm BH}$ & $\log L_{\rm R,VLA}$ & 0.74 & $5.79\times10^{-3}$ \\
		$\log L_{\rm X,14-195keV}$ & $\log M_{\rm BH}$ & $\log L_{\rm R,VLBI}$ & 0.74 & 0.07 \\
		\hline
	\end{tabular}
	\parbox[]{\columnwidth}{
	{\it Note.} 
	Column (1)-(3): Variable $X$, $Y$ and $Z$, respectively. Correlation between variables $X$ and $Y$ is studied, taking into account the mutual correlation of $X$ and $Y$ with the third variable $Z$. 
	Column (4)-(5): the partial Kendall's correlation coefficient $\tau$, and the probability for accepting the null hypothesis that there is no correlation between $X$ and $Y$. 
	}
	\end{center}
\end{table}

We plot X-ray luminosity versus radio luminosity in Figure~\ref{xvr}. 
We calculate the Spearman's rank correlation coefficient $\rho$ and the probability of null hypothesis (i.e., no correlations), $P_{\rm null}$, 
to test whether there is significant correlations between the two variables, 
using the X-ray luminosities either in $2-10\rm\,keV$ or in $14-195\rm\,keV$, 
and the radio luminosities calculated from either VLA or VLBI observations. 
However, all the tests give $P_{\rm null}>5\%$, 
which means that there is no significant correlation between the radio luminosity and 
X-ray luminosity.

By analyzing a large sample of AGNs, it was found that the radio luminosity depends on both the X-ray luminosity and black hole mass 
\citep[e.g.,][]{2003merloni}. 
In the current paper, more generally,  we explore the mutual relation between  $L_{\rm R}$,  $L_{\rm X}$, $M_{\rm BH}$ and 
nuclear monochromatic luminosity $L_{5100}$ by using the partial Kendall's $\tau$ correlation test, which can be used to eliminate the 
effect of a third variable when assessing the correlation between two variables 
\citep{1996MNRAS.278..919A}.

Firstly, we calculate the partial correlation coefficients $\tau$ and the probabilities for null hypothesis $P_{\rm null}$ 
between either two of the radio luminosity $L_{\rm R}$, X-ray luminosity $L_{\rm X}$ and optical luminosity $L_{5100}$, 
given that the black hole mass $M_{\rm BH}$ is the third variable. 
Only a weak correlation is revealed by the test between $L_{\rm R,VLA}$ and $L_{5100}$ with probability of $P_{\rm null}=4\%$. 
Then we test the correlations between $L_{\rm X}$ and $L_{\rm R}$ given that $L_{5100}$ is the third variable. 
We don't find significant correlations between $L_{\rm R}$ and $L_{\rm X}$ in any cases so far. 
Finally, we test if there is correlation between $L_{\rm R}$ (or $L_{\rm X}$) and $M_{\rm BH}$, 
given that $L_{5100}$ or $L_{\rm X}$ (or $L_{\rm R}$) is the third variable. 
No significant correlation has been found between $L_{\rm R}$ and $M_{\rm BH}$. 
However, 
both $2-10\rm\,keV$ and $14-195\rm\,keV$ luminosity significantly  correlate with $M_{\rm BH}$ with $P_{\rm null}<1\%$.

\begin{figure*}
\centering
	\begin{tabular}{cc}
		\includegraphics[width=0.9\columnwidth]{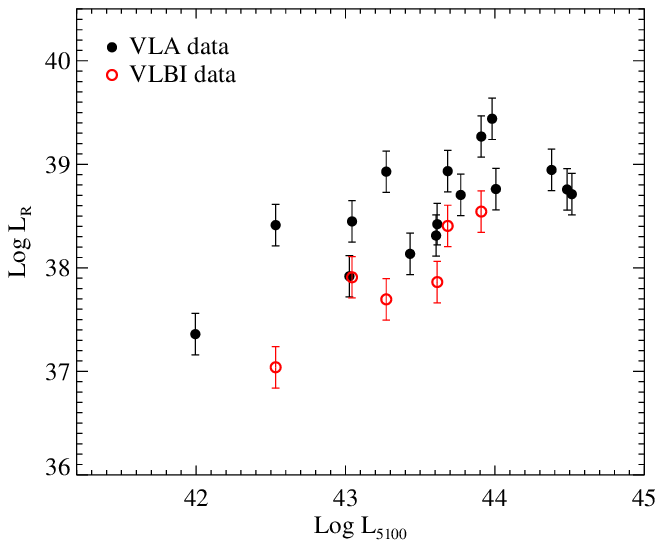} &
		\includegraphics[width=0.9\columnwidth]{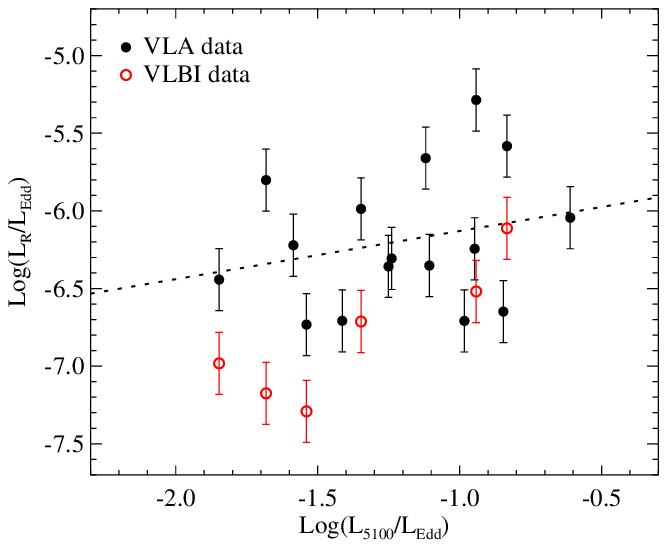} \\
	\end{tabular}
	\caption{
	{\it Left panel:} 
	The radio luminosity $L_{\rm R}$ versus the monochromatic nuclear optical luminosity $L_{5100}$. 
	{\it Right panel:} 
	$L_{\rm R}/L_{\rm Edd}$ versus $L_{5100}/L_{\rm Edd}$. 
	The error bars represent typical uncertainties $\sigma_{\rm R}=0.2$ of the radio luminosities. 
	The black dotted line in the right panel is the best linear fit to the relation of $L_{\rm R}/L_{\rm Edd}$ and $L_{5100}/L_{\rm Edd}$ using VLA data. 
	}
\label{r_optical}
\end{figure*}

In the left panel of Figure~\ref{r_optical}, 
we plot $L_{\rm R}$ versus $L_{5100}$ using VLA and VLBI data respectively. 
It is clear that $L_{\rm R}$ increases with increasing $L_{5100}$. 
Then, 
as suggested by the partial correlation test (the fifth row of table~\ref{cor_tab}), 
we plot the Eddington-scaled radio luminosity $L_{\rm R}/L_{\rm Edd}$ versus the Eddington-scaled optical luminosity $L_{5100}/L_{\rm Edd}$ in the right panel of Figure~\ref{r_optical}. 
By adopting a typical uncertainty $\sigma_{\rm R}=0.2$ dex for the radio luminosity \citep[][]{2001ApJ...555..650H}, 
the best-fitted linear relation  
using VLA data gives
\begin{equation}
	\log\left(\frac{L_{\rm R,VLA}}{L_{\rm Edd}}\right)=(0.31\pm0.33)\log\left(\frac{L_{5100}}{L_{\rm Edd}}\right)-(5.82\pm0.41). 
\end{equation}

\begin{figure*}
\centering
	\begin{tabular}{cc}
		\includegraphics[width=0.9\columnwidth]{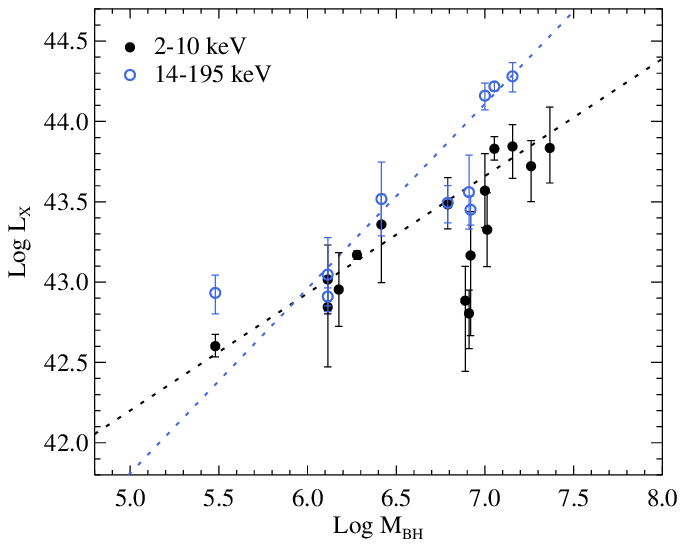} &
		\includegraphics[width=0.9\columnwidth]{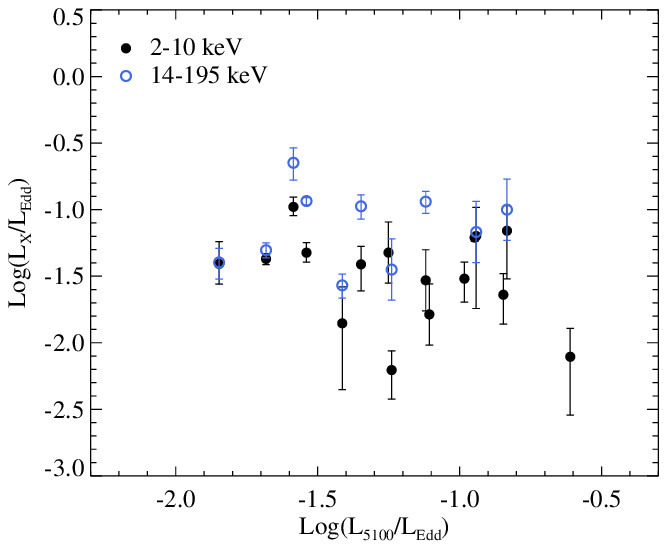} \\
	\end{tabular}
	\caption{
	{\it Left panel:} 
	The black filled circles and black dotted line represent respectively the $\log M_{\rm BH}$ versus $\log L_{\rm X}$ relation using $2-10\rm\,keV$ luminosity and the best linear fit, 
	while the blue open circles and blue dotted line represent respectively the same relation using $14-195\rm\,keV$ luminosity and the best linear fit. 
	{\it Right panel:} 
	The Eddington-scaled optical luminosity $L_{5100}/L_{\rm Edd}$ versus the Eddington-scaled X-ray luminosity $L_{\rm X}/L_{\rm Edd}$ in $2-10\rm\,keV$ and $14-195\rm\,keV$, respectively. 
	}
\label{x_mass}
\end{figure*}

In the left panel of Figure~\ref{x_mass}, 
we plot $M_{\rm BH}$ versus $L_{\rm X}$ using $2-10\rm\,keV$ luminosities and $14-195\rm\,keV$ luminosities respectively. 
The best-fitted linear relations of $M_{\rm BH}$-$L_{\rm X}$ are 
\begin{equation}
	\log L_{\rm X,2-10\,keV}=(0.73\pm0.10)\log M_{\rm BH} +(38.55\pm0.60)
\end{equation}
and 
\begin{equation}
	\log L_{\rm X,14-195\,keV}=(1.15\pm0.13)\log M_{\rm BH} +(36.06\pm0.91). 
\end{equation}
In the right panel of Figure~\ref{x_mass}, 
as suggested by the partial correlation test (the seventh and eighth row of table~\ref{cor_tab}), 
we plot the Eddington-scaled X-ray luminosity $L_{\rm X}/L_{\rm Edd}$ versus the  Eddington-scaled optical luminosity $L_{5100}/L_{\rm Edd}$. 
No significant correlation between  $L_{\rm X}/L_{\rm Edd}$ and $L_{5100}/L_{\rm Edd}$ is found.

\section{Discussion}
\label{discussion}

We collect the core radio luminosities calculated from high-resolution radio observations, 
as well as the X-ray luminosities for 16 nearby NLS1s with Eddington ratios $\gtrsim0.1$. 
The high-resolution radio images avoid the host galaxy contamination as much as possible. 
Partial Kendall's $\tau$ correlation test is used to explore the mutual relations between the radio luminosity, X-ray luminosity, 
nuclear optical luminosity and the black hole mass. 
As shown in Section~\ref{correlations}, 
we don't find any significant correlations between the X-ray luminosity $L_{\rm X}$, 
either in $2-10\rm\,keV$ or $14-195\rm\,keV$, 
and the radio luminosity $L_{\rm R}$, 
either using VLA data measured on scales of $\lesssim1\rm\,kpc$ 
or VLBI data measured on scales of a few pc. 
However, 
we find that there is a positive correlation between  $L_{\rm X}$ and $M_{\rm BH}$. 
Meanwhile, we find that there is no correlation between $L_{\rm X}/L_{\rm Edd}$ and $L_{5100}/L_{\rm Edd}$ (right panel of Figure~\ref{x_mass}), 
which are suggested to be understood in the framework of slim disc scenario as follows.

The slim disc model has been proposed to explain the observational features of NLS1s in the past few decades 
\citep[e.g.,][]{1988slim, 1999ApJ...516..420W, 2000slim, 2003A&A...398..927W, 2013PhRvL.110h1301W}. 
In this model, 
when the accretion rate approaches or exceeds the Eddington limit, 
the disc within a certain radius $R_{\rm trap}$ becomes optically too thick 
that the time scale of the photon diffusion to the disc surface is longer than that of the viscous time scale of the accretion disc. 
In this case, most of the photons within this radius will be trapped in 
the disc until advected into the black hole rather than being radiated out. 
The radius of the photon trapping region is dependent on the accretion rate as $R_{\rm trap}\propto\dot{M}_{\rm disc}$ 
\citep{1999ApJ...516..420W, 2013PhRvL.110h1301W}. 
As $\dot{M}_{\rm disc}$ increases,  the photon trapping region gets larger. 
The total luminosity of the slim disc is proportional logarithmically to $\dot{M}_{\rm disc}$
and eventually get saturated with increasing $\dot{M}_{\rm disc}$. 
Then the total luminosity is independent on $\dot{M}_{\rm disc}$ and only depends on the black hole mass $M_{\rm BH}$ \citep[][]{2013PhRvL.110h1301W}.

The hard X-ray emission above $2\rm\,keV$ in bright Seyfert galaxies and radio quiet quasars is believed to be from a so-called 
disc-corona system, which should be very similar in the case of NLS1s with a form of slim disc-corona system \citep[e.g.,][]{2004ApJ...614..101C, 2017MNRAS.468.3663J}. 
The emission from the disc mainly contributes to the optical/UV band, part of which will be Compton up-scattered in the hot corona,
producing the hard X-ray emissions \citep[e.g.,][]{1991ApJ...380L..51H, 1993ApJ...413..507H, 1995ApJ...449L..13S}. 
Simply, the Compton cooling rate is determined by the following expressions \citep[][]{1986rpa..book.....R}, 
\begin{equation}
	q_{\rm cooling}=\frac{4kT_{e}}{m_{e}c^{2}}n_{e}\sigma_{\rm T}cu, 
\end{equation}
where 
$u$ is the seed photon energy density, 
$T_{e}$ is the electron temperature, 
$\sigma_{\rm T}$ is the Thomson scattering cross section, 
$n_{e}$ is the electron density and 
$m_{e}$ is the electron mass. 
If the NLS1s in our sample are indeed powered by the slim disc-corona accretion, we will have a  general consequence as follows. 
When the accretion rate gets near and above the Eddington limit as in the NLS1s studied here (Column 15 of Table~\ref{rqnls1_sample}), 
the emission from the accretion disc will be saturated and independent on the accretion rate within the photon trapping radius $R_{\rm trap}$. 
Observationally, since the corona is very compact, 
located within less than $\sim20$ gravitational radii of the black hole 
\citep[e.g.,][]{2013ApJ...769L...7R},
the covering factor of the accretion disc at the region beyond a few tens of gravitational radii as seen from the corona is very small. 
In this case, the seed photons to be intercepted and scattered in the corona will be mainly from the very innermost region of the accretion disc. 
So, in the slim disc scenario, although the luminosity of  the accretion disc beyond $R_{\rm trap}$ do not get saturated, 
however, due to the relatively small contribution of the seed photon from this region, 
the seed photon energy density $u$ for the inverse Compton scattering in the corona will not change much 
with increasing $\dot{M}_{\rm disc}$. 
Meanwhile, 
if the property of the corona, such as the electron temperature and the optical depth also don't change, 
theoretically, 
the produced hard X-ray emission from the corona will not change much. 
Then the overall hard X-ray emission will be only dependent on the black hole mass, 
which is just the correlation we found between $L_{\rm X}$ and $M_{\rm BH}$ (left panel of Figure~\ref{x_mass}).

Physically,  in the standard disc-corona case, the relative strength between the disc and the corona is determined by the relative mass accretion rate in the disc and corona. 
Generally, the radiation from the disc, $L_{\rm disc}$, 
increases with an increase of the mass accretion rate in the disc, $\dot{M}_{\rm disc}$, 
which will make the matter in the corona collapse due to the strong Compton cooling processes, 
decreasing the relative strength of the emission from the corona. 
In the slim disc-corona case, however, 
an increase of $\dot{M}_{\rm disc}$ does not increase the seed photon luminosity, 
which will not change the coronal luminosity, 
then consequently lead to the independence of the coronal luminosity on the accretion rate.

In summary, 
in the slim disc-corona scenario, 
with an increase of $\dot{M}_{\rm disc}$, 
the disc luminosity increases, leading to an increase of $L_{5100}$. 
However, the X-ray luminosity nearly does not change. 
This is strongly supported by our finding in the current paper, i.e., there is no observed correlation between $L_{\rm X}/L_{\rm Edd}$ and $L_{5100}/L_{\rm Edd}$ (right panel of Figure~\ref{x_mass}), 
meanwhile, there is a positive correlation between $L_{\rm X}$ and $M_{\rm BH}$ (left panel of Figure~\ref{x_mass}). 
We note that, 
in some numerical simulations of supercritical accretion flows, 
the fully saturated luminosity of the accretion disc occurs at Eddington ratio $\sim3$ \citep[e.g.,][]{2005ApJ...628..368O}, 
which is a little higher than that of the sources in our sample, i.e., $0.14-2.4$ (Table~\ref{rqnls1_sample}). 
Obviously, 
when the Eddington ratio is in the range of $0.14-2.4$, 
the disc is partly saturated, 
which can intrinsically predict our findings in Figure~\ref{x_mass}. 

The relativistic jet  has been widely observed in both AGNs and BHBs. 
So far, several models have been proposed for the jet formation, 
such as the Blandford-Znajek (BZ) process 
and Blandford-Payne (BP) process.
In the BZ process, the jet is driven by extracting the rotational energy of the black hole via a large-scale magnetic field \citep{BZ}.
In the BP process, the magnetic fields thread the accretion disc, extracting the  rotational energy of the accretion disc to drive the jet \citep{BP}. 
Some state-of-the-art numerical simulations of super-Eddington accretion around a supermassive black hole have been done for the formation of jet, 
in which the jet is driven by radiation pressure 
\citep[e.g.,][]{2014MNRAS.437.2744T, 2015MNRAS.453.3213S}.
We should note that, although a lot of important progresses have been made for the formation of the jet, 
we don't fully understand the physical mechanism for launching the jet yet \citep[e.g.,][]{2017SSRv..207....5R}

As we know, a strong correlation  between the radio luminosity and X-ray luminosity has been found both in AGNs and BHBs \citep[e.g.,][]{2003merloni}.
The advection dominated accretion flow (ADAF) plus jet model was proposed to explain the correlation, 
i.e., $L_{\rm R}\propto L_{\rm X}^{0.5-0.7}$ in LLAGNs and in the low/hard spectral state of BHBs for $L_{\rm X}/L_{\rm Edd} \lesssim 10^{-3}$ 
\citep[e.g.,][]{2005ApJ...629..408Y, 2016MNRAS.456.4377X}. 
The disc-corona plus jet was proposed to  explain the correlation, i.e., 
$L_{\rm R}\propto L_{\rm X}^{1.4}$ for $L_{\rm X}/L_{\rm Edd} \gtrsim 10^{-3}$ 
in luminous AGNs and the luminous low/hard spectral state of BHBs 
\citep[e.g.,][]{2014ApJ...787L..20D, 2015MNRAS.448.1099Q}. 
However, for the NLS1s in current paper, we don't find a significant correlation between the radio luminosity and X-ray luminosity, which may 
imply that the physics for the coupling between the accretion (especially the corona) and jet in NLS1s is completely different from other types 
of AGNs and the low/hard spectral state of BHBs. 
Due to the high Eddington ratios,  NLS1s are argued to be the  scale-up version of BHBs in their high/soft state or very high state 
\citep[e.g.,][]{2006MNRAS.372..401A}.
However, the relation between the radio and X-ray luminosities in the high/soft state or very high state of BHBs is not clear now \citep[e.g.,][]{2016MNRAS.463..628R}. 
Thus, we do not compare their radio-X-ray relations here.

In \citet[][]{2003merloni}, 
different kinds of black hole sources were included to study the mutual dependencies of $L_{\rm R}$, $L_{\rm X}$ and $M_{\rm BH}$, 
including the BHBs in the low/hard spectral state, 
and different types of AGNs (LLAGNs, Seyfert galaxies and quasars). 
They have looked for the partial correlations between $L_{\rm R}$ and $L_{\rm X}$ by taking $M_{\rm BH}$ as the third variable. 
Then they have also looked for partial correlations between $L_{\rm R}$ ($L_{\rm X}$) and $M_{\rm BH}$ by taking $L_{\rm X}$ ($L_{\rm R}$) as the third variable. 
In their work, $L_{\rm R}$ is strongly correlated with both $M_{\rm BH}$ and $L_{\rm X}$, 
and, in turn, $L_{\rm X}$ is correlated with both $M_{\rm BH}$ and $L_{\rm R}$ when both AGNs and BHBs are considered, 
which then leads to the discovery of the so-called `fundamental plane'. 
In our work, 
we only focus on a special subclass of AGNs, i.e., NLS1s, with relatively higher Eddington ratios. 
We don't find partial correlations between the radio luminosity and X-ray luminosity, 
and between the radio luminosity and black hole mass (Table~\ref{cor_tab}). 
So the determination of a plane is impossible here. 
However, 
we find a significant partial correlation between the $L_{\rm X}$ and $M_{\rm BH}$ for NLS1s 
by taking $L_{\rm R}$ or $L_{5100}$ as the third variable. 
This is in contrast with \citet[][]{2003merloni}, 
in which $L_{\rm X}$ is found to be independent on $M_{\rm BH}$ when only AGNs are included (see their Table 2). 
This also implies that the coupling between the accretion and jet in the NLS1s is indeed different from other types of AGNs.

But we should note that, 
in the current paper, the number of the sources in the sample is small, which may make the statistical  properties uncertain. 
A larger well-defined and uniformly selected sample with high resolution radio images and X-ray observations is needed to confirm our conclusions more robustly in the not far future. 

\section*{Acknowledgements}

We are grateful to anonymous referee for the helpful comments, 
which substantially improve the quality of this paper. 
We thank 
W. M. Yuan, B. F. Liu, Z. Liu, S. Komossa and H. Y. Zhou 
for helpful discussions. 
This work has made use of the NED which is operated by the Jet Propulsion Laboratory, California Institute of Technology, under contract with the National Aeronautics and Space Administration.
S. Yao thanks the supported by KIAA-CAS fellowship. 
E. L. Qiao thanks the supports by the NSFC grant No. 11773037. 
X. B. Wu thanks the supports by the Ministry of Science and Technology of China under grant 2016YFA0400703, the NSFC grants No.11373008 and 11533001, and the National Key Basic Research Program of China 2014CB845700.

\bibliographystyle{mnras}
\bibliography{references}

\bsp	
\label{lastpage}
\end{document}